# WEAPONIZATION OF CONSCIENCE IN CYBERCRIME AND ONLINE FRAUD: A NOVEL SYSTEMS THEORY

Michelle Espinoza
Marymount University, Arlington, USA
M0e73021@marymount.edu

*Abstract:* This article introduces the concept of weaponization of conscience as a complex system and tactic employed by fraudsters to camouflage their activity, coerce others, or to deceive their victims. This study adopts a conceptual approach, drawing from the theoretical underpinnings of military propaganda and psychological operations doctrines and adapting them to serve as a lens through which to understand and defend against weaponization of conscience.

## 1. INTRODUCTION

When a problem is growing exponentially, such as cybercrime, it is because a positive feedback loop is dominating the system at that point in time. Exponential growth is limited by the carrying capacity of the system, and to regulate, slow, or reverse cybercrime's growth, one must identify the positive feedback loop and then counter it with a negative feedback loop or break the positive feedback loop entirely. The problem's growth may be within a department, an organization, and/or society. Systems are akin to matryoshka (nesting) dolls, each subsystem embeds within and interacts with a larger system. This is why cybercrime and online fraud require examination through the lens of complexity science (Espinoza, 2024). Constructing a solution within an organization or at a societal level requires understanding how it works, why it works, and/or why it does not work.

This research introduces the concept of *weaponization of conscience* as a complex system and tactic employed by fraudsters to camouflage their activity, coerce others, or to deceive their victims. Weaponization of conscience can be used alone or in combination with other tactics such as weaponization of bias and weaponization of silence to synergize results and extend the effectiveness of their modi operandi. Additionally, weaponization of conscience can be employed against coerced victims to strengthen their attack against the target. Weaponization of conscience is a complex subsystem within the larger complex system of social engineering. This article contributes to the ongoing discourse on theoretical innovation and the advancement of knowledge in cybercrime and online fraud detection from a system dynamics perspective.

Hardin's seminal article on the Tragedy of the Commons posits that to appeal to an individual's conscience is to intentionally opt for the extinction of all conscience in the long run (Hardin, 1968). Hardin refers to the conscience as pathogenic and likens the conjuring of conscience to using a biological weapon. Hardin's arguments are rooted in Darwin's (1859) theory on survival of the fittest, which posits that organisms best adjusted to their environment are the most successful in surviving and reproducing. Put simply, to act with empathy, compassion, altruism is to sentence oneself to death in a world of barbarians. But without empathy and compassion for one another, what more are we but barbarians? Defending against the weaponization of conscience, requires striking the balance between self-awareness and compassion or pro-social behaviors.

The human mind is considered a new domain of operations across Western, Chinese, and Romanian information operation doctrines (Neculcea & Răpan, 2022). Deceptive practices are reaching new levels of realism and complexity due to technological advances (Masakowski & Blatny, 2023; Onushi, 2023). There exists a vast body of research on decision-making processes within highly specific contexts. In contrast, the realm of decision-making in information warfare is a disputable and neglected area of investigation (Maksymenko & Derkach, 2023). The deceptive practices used by fraudsters and scammers bear a striking resemblance to the dynamics at play in the landscape of information warfare where misinformation, deceit, manipulation, and other psychological tactics are employed to achieve strategic objectives.

## 2. PROBLEM STATEMENT

Fraud and other crime are not new, however digitization and the internet transformed the threat landscape so that crimes once requiring proximity of the offender to their victim, can now be carried out across the world with the click of a button (Cross, 2020). Transnational collusion, jurisdictional issues, and the exponential growth of cybercrime present significant implications for law enforcement as it relates to their ability to investigate, arrest, and prosecute offenders (Cross, 2020; Kshetri, 2017; Salih & Dabagh, 2023). Consequently, many municipalities have insufficient funding and resources to protect their citizenry from victimization by fraudsters (Darii & Meleca, 2022). Understanding how and why fraudsters' modi operandi succeed is essential to constructing an effective defense (Rutherford, 2017). 95% of cyber incidents are human-enabled (Nobles, 2018). Modeling decisions and decision interactions as complex systems enables practitioners to incorporate situational or domain-specific details with decision theories to pinpoint vulnerabilities in historical and prospective cyber defense strategies. This article:

1. Defines the weaponization of conscience as a complex system.
2. Explores how the principles of system dynamics are applicable to the weaponization of conscience.

3. Provides examples of how weaponization of conscience manifests, why it may be successful, and presents strategic recommendations for fortifying defenses against such tactics.

## 3. METHOD

This study adopts a conceptual approach, drawing from the theoretical underpinnings of military propaganda and psychological operations doctrines and adapting them to serve as a lens through which to understand and conceptualize how cognitive strategies, specifically the weaponization of conscience, is employed by fraudsters in camouflaging their activities and deceiving victims. Writing theory is challenging, partly due to the lack of straightforward formulas or templates (Cornelissen, 2017; Ragins, 2012). Theory adaptation revises extant knowledge by proposing a novel perspective on an extant conceptualization whereas typology aims to develop a categorization that explains the fuzzy nature of many subjects by logically and causally combining different constructs into a more nuanced understanding of a phenomenon or concept (Cornelissen, 2017; Jaakkola, 2020). This paper uses a hybrid of theory adaptation and typology. Building on the legal definitions and frameworks of conscience laws in the United States and using Greene's (2009) dual-process theory, this researcher provides a base system model template which practitioners can use to visualize decision processes and interactions. By incorporating the critical features of psychological operations, the principles of system dynamics, and using extant cognition and behavior theories from multiple disciplines, this article then provides practical examples of how weaponization of conscience might manifest and presents strategic recommendations to defend against it.

## 4. CONSCIENCE LAWS

The history of conscience laws in the United States reflects a complex interplay between legal, ethical, and religious considerations. Conscience laws, often referred to as conscience clauses or accommodation laws, are designed to protect individuals and entities from being compelled to participate in activities that conflict with their deeply held moral or religious beliefs. In summary, the history of conscience laws in the U.S. reflects a dynamic evolution, driven by changing societal values, legal interpretations, and ongoing debates about the balance between religious freedom and the rights of individuals (Health and Human Services Department, 2024; Office for Civil Rights (OCR), 2010) . The purpose behind incorporating references to conscience laws within this theory adaptation is to underscore the notion that conscience transcends mere philosophical abstraction. By grounding the discussion on conscience as defined in within a legal context, we can focus on the mechanics of the theory.

For this framework, conscience is defined as an intrinsic motivational force guided by a person's moral beliefs. Those moral beliefs may or may not be rooted in religion and a person's moral beliefs can change or evolve. Moral beliefs or decisions can also be situational or relative. One need only ponder a variant of the trolley problem to realize how moral dilemmas can scramble our understanding of decision-theory.

## 5. PSYOPS

Propaganda and other variants of psychological manipulation have been used since the dawn of time. The US Department of Defense updated its doctrine on psychological operations and clarified that the purpose of PSYOPS is to convey selected information, indicators to foreign audiences to influence their emotions, motives, objective reasoning, and ultimately the behavior of foreign governments, organizations, groups, and individuals favorable to the originator's objectives (US Department of Defense, 2010). The term PSYOPS has been used interchangeably with information warfare, psychological warfare, and information operations at different points in history, but the overarching objective has always been to influence the target's behaviors through information and emotion (Hwang & Rosen, 2019; Linebarger, 1950). According to Pop (2017), all critical PSYOPS have three core objectives: acquiring knowledge, thwarting the enemy's attempts at the acquisition of knowledge, and misleading the enemy to build a database containing inefficient and useless information. Malyuk (2022) argues that the most important elements in information operations are informational and psychological suppression of the enemy, but also acknowledges that if citizenry know how to recognize the tactics of information warfare it weakens and minimize their effectiveness (Malyuk, 2022). Fraudsters use the weaponization of conscience to influence their victims' behaviors through information and emotion.

## 6. COMPLEX SYSTEMS

Complexity science is the study of the relationship between systems. At its inception in the 1950s, system dynamics focused primarily on corporate and industrial problems such as production cycles, and supply chains. It relied heavily on hand-drawn simulations and calculations before computer modeling was available. Today, the principles of system dynamics are applied across various disciplines and industries including public policy, healthcare, and social systems (Gentili, 2021). The transferable principles of dynamic systems and amenability to qualitative manual model construction make system dynamics well-suited for modeling the weaponization of conscience. Complex systems are highly interdependent, decentralized, self-organized, adaptive, and sensitive to feedback loops (Forrester, 2003). Complex systems may also exhibit the butterfly effect, borrowed from chaos theory, where a small change in the system exhibits an unexpectedly amplified effect in the output (Samoilenko & Osei-Bryson, 2007). According to the system dynamics perspective, every decision is made in the context of a feedback loop, a circular flow of information or effects that cause the system to regulate or adjust its behavior based on the information returned in each cycle (Forrester, 2003). Complex systems obey the laws of conservation and accumulation. The law of conservation states that energy cannot be created or destroyed, it can only be transferred or transformed. A key distinction between linear and complex systems is that the behavior of a linear system can be understood by examining its individual parts whereas a complex system is highly interconnected, and the system behavior emerges from the interactions between its parts. Weaponization of conscience is an example of a complex system.

## 7. DUAL PROCESS OF MORAL JUDGEMENT

Dual-process theory of moral judgement asserts that moral decisions are the result of either one of two distinct mental processes, intuitive or conscious (Greene, 2009). Greene's dual-process model of moral judgement posits that intuitive judgements are fast, emotionally-driven, and primarily deontological (rule-based) whereas conscious decisions are deliberative and characteristically utilitarian (Greene, 2009).

For simplicity, the subcomponents of the conscience system are described in binary terms, but it is important to note that each subcomponent is dynamic, interacts with the other subcomponents, and rests on a continuous scale as depicted in (Figure 1). Moral beliefs shift, emotions shift and vary in intensity. Additionally, this researcher posits that no decisions or actions can be taken without passing through each phase depicted in (Figure 1), and the subsystem is activated by an external event or interaction not depicted.

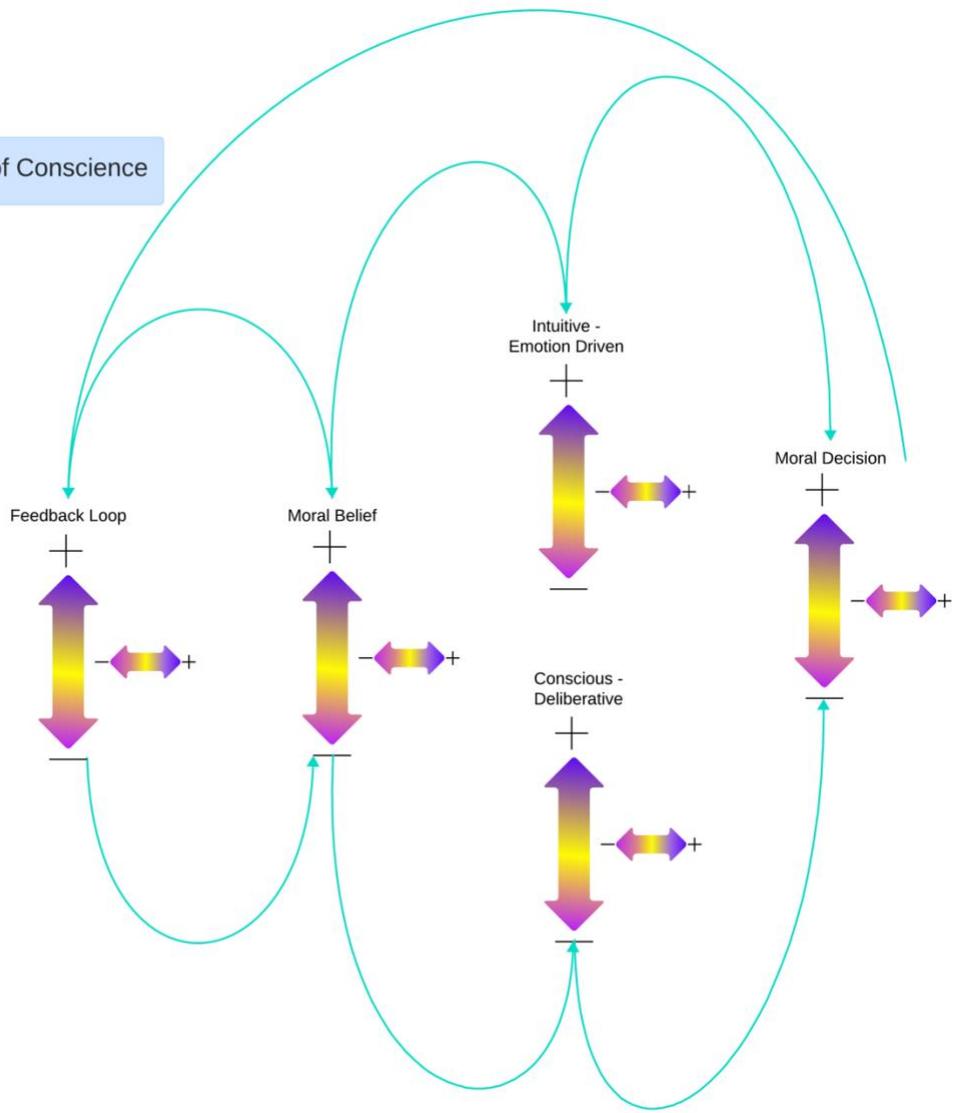

*Figure 1Weaponization of conscience. **Note**: Author's own work. No copyright attribution required.*

## 8. APPLICATION

### 8.1 Scam Call Example

Financial exploitation from scam calls is one of several growing variants of cybercrime and online fraud costing victims billions each year (Doyle, 2022). The scammer may call their victim posing as a loved one in distress needing money or ransom. A higher level of distress is likely to intensify the victim's level of fear and urgency. Their decision to act, will skew heavily towards intuitive cognitive processes. Now, assume that the scammer(s) call and pose as two loved ones in distress needing money or ransom, but the victim can only afford the ransom for one. The dilemma shifts the victim's moral decision processes more heavily towards conscious/deliberative processes, but the emotional intensity will be [arguably] higher than the intensity experienced in the single-hostage scenario. Increased fear levels typically slow decision-making and increase deliberative processes (Wake et al., 2020; Wang et al., 2023). The victim's decision processes are strained by the additional cognitive load and weakened further by the intensification of emotion. When moral dilemmas are personal to the decision-maker, studies have demonstrated that utilitarianism is reduced (Starcke & Brand, 2012). Additional variables that might influence how the victim responds include the relationship to the hostage, the intensity of threat, the victim's gender, age, or experience with similar scams.

To counter this double-load effect, one solution may be to hang up the phone and call the loved one directly. Another solution may be to create a pre-arranged code word to authenticate loved ones' identities in these situations. This researcher does not seek to mechanize or diminish the real anguish experienced by millions of victims across the world. Rather, the example depicted in (Figure 2) is intended to illustrate the concept and enable readers to think through cognitive processes from a systems perspective to identify vulnerabilities and tactics used by fraudsters.

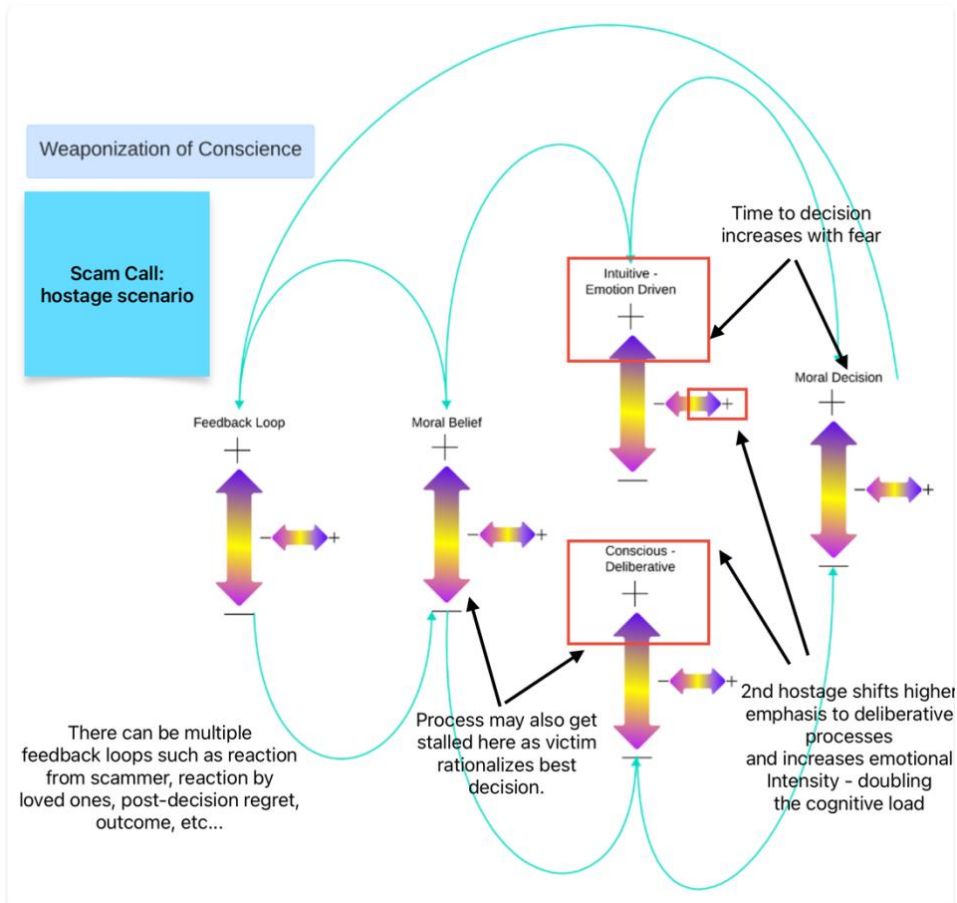

*Figure 2 Weaponization of Conscience - Scam Call Scenario. Note: Author's own work. Attribution not required.*

## 8.2 Combination Effect – Identity Theft and Phishing

Voter registration lists are public record in several countries. Knowing a victim's voter party-affiliations, particularly in hyper-partisan countries, enables scammers to estimate their target's belief system and causes or passions that may evoke intense emotional responses. Using the moral belief 'dial' from the weaponization of conscience in (Figure 3), assume that the target's intensity of belief in a specific cause is in the middle of the spectrum. Combining McGuire's (1964) inoculation theory, whereby initial weak counter-attitudinal messages lead to stronger resistance against subsequent counter-attitudinal messages with valence-framing, where framing a person's belief as negative or oppositional leads to firmer entrenchment in that belief (Bizer et al., 2011), fraudsters can increase the intensity of the moral belief dial and potentially the intensity of the intuitive dial before presenting the victim with a phishing site seeking recurring donations in support of a cause likely to appeal to the target. The priming effect of manipulating the victim's conscience system before soliciting donations to a fictitious organization may yield higher donation amounts than if the priming had not been done beforehand. Defending against this weaponization of conscience requires potential victims to be conscious of the reaction that messaging evokes and to consider the source and motives of the messenger.

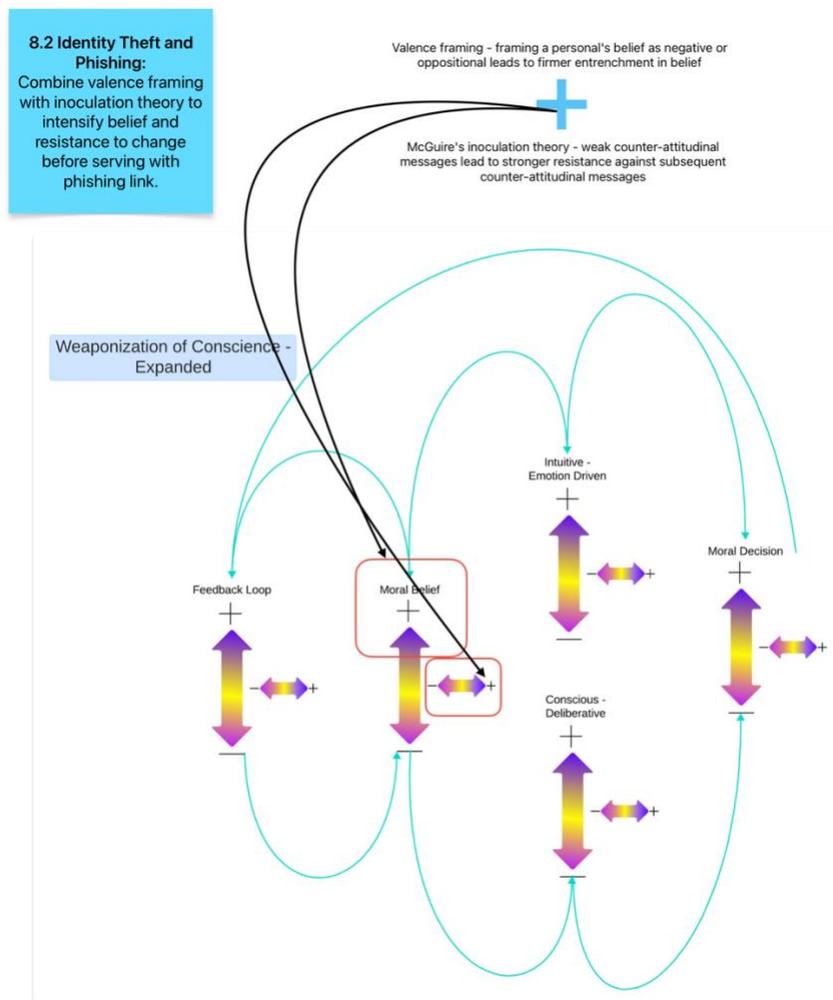

*Figure 3 Weaponization of Conscience - Phishing Link Example. Note: Author's own work. Attribution not required.*

## 9. CONCLUSION

This article introduced the concept of *weaponization of conscience* as a complex system and tactic used by fraudsters to camouflage their activity, coerce others, or to deceive their victims. Weaponization of conscience is a complex subsystem within the larger complex system of social engineering. This framework defines conscience as an intrinsic motivational force guided by a person's moral beliefs. Those moral beliefs may or may not be rooted in religion and a person's moral beliefs can change or evolve. Moral beliefs or decisions can also be situational or relative.

The decision system is always activated by an event or stimulus from outside the conscience subsystem such as when a user is presented with a phishing link or receives a scam call. The starting point (state) of the conscience system is influenced by multiple factors such as preceding events, mood, time of day, priming events. It is not unusual for multiple feedback loops to manifest in the subsystem and sometimes these feedback loops will have different delay times. Modeling decisions and decision interactions as complex systems enables practitioners to incorporate situational or domain-specific details with decision theories to pinpoint vulnerabilities in historical and prospective cyber defense strategies.

Hardin's argument that the conscience weakens humanity may be true, but just as we developed bulletproof vests because a person cannot survive without a beating heart, so too we must develop defenses against the weaponization of conscience. Without conscience, humanity cannot survive.